\documentstyle[preprint]{ptptex}

\newcommand{\bra}[1]{\langle {#1} |}     
\newcommand{\ket}[1]{| {#1} \rangle}     
\newcommand{\wtilde}[1]{\widetilde{#1}} 

\title{
Parametric Resonance versus Forced Oscillation in \\
Time-Evolution of Quantum Meson Fields
}

\author{
Yasuhiko {\sc Tsue}
}

\inst{
Physics Division, Faculty of Science, Kochi University, Kochi 780-8520, 
Japan
}

\recdate{
}

\abst{
The time-evolution of quantum meson fields in O(4)-linear sigma model 
is treated approximately. 
It is shown that the amplification of the amplitudes of 
pion modes with low momenta occurs by means of both the parametric resonance 
and the forced oscillation. 
}

\begin{document}

\maketitle


It is interesting to study the dynamics of the chiral phase transition 
in the context of relativistic heavy-ion collisions, such as the problem 
of the formation of a disoriented chiral condensate. One of the important 
theoretical aspects is to investigate the time-evolution of the order 
parameter of the chiral phase transition. As for the fluctuation modes 
around the order parameter, the amplification of the amplitudes 
must occur accompanying with the relaxation of the chiral order parameter. 
Recently, the present author has investigated the time-evolution 
of a collective meson field in the context of the dynamical chiral phase 
transition.\cite{TKI} 
It has been seen that the amplitudes of quantum fluctuations 
with low momenta around 
the mean field configuration have been amplified with the direction 
of pion modes in the $O(4)$-linear sigma model. 
As is indicated by many authors, this phenomena may be understood in 
terms of a parametric amplification.\cite{MM,HM,M,I}
However, there may be another mechanism to amplify the fluctuation modes. 
Actually, it is seen that the amplitude of the pion mode with 
higher momentum without ${\mib k}=0$ can be amplified in the previous 
paper.\cite{TKI}
In this paper, it is pointed out that there is another mechanism to amplify 
the quantum fluctuation modes through the chiral phase transition, namely, 
that a forced oscillation works as well as a parametric resonance.

Before treating the $O(4)$-linear sigma model, it is instructive to 
recapitulate 
the ingredients of the parametric resonance and the forced oscillation. 
Let us consider the following equation of motion for $x(t)$ : 
\begin{equation}\label{1}
{\ddot x}+\omega_0^2(1-h\cos \gamma t)x = 0 \ .
\end{equation}
The parametric resonance occurs around $\gamma=2\omega_0/m\ 
(m=1, 2, \cdots)$.\cite{LL} 
For $m=1$, the resonance region of the frequency $\gamma$ is given as 
\begin{eqnarray}\label{2}
& &-{h\omega}/{2} < \epsilon < {h\omega}/{2} \ ,\qquad
\hbox{\rm for}\quad \gamma=2\omega_0 + \epsilon \ ,\quad \epsilon \ll \omega_0 
\ , 
\end{eqnarray}
where we treated $h$ as a small parameter, $h\ll 1$. Similarly, for $m=2$, 
we obtain 
\begin{eqnarray}\label{3}
-({5}/{24})\cdot h^2\omega_0 < \epsilon < ({1}/{24}) 
\cdot h^2\omega_0 \ ,\qquad
\hbox{\rm for}\quad \gamma=\omega_0 + \epsilon \ ,\quad \epsilon \ll \omega_0 
\ . 
\end{eqnarray}
In general, for given integer $m$, the unstable solution of $x(t)$ is 
obtained as the form 
$x(t)\propto e^{st}\times$(oscillation and constant parts), 
where $s$ is of the order of $h^m$. Namely, the amplification becomes 
slower in time as the integer $m$ becomes larger. 
The resonance region also becomes 
narrow with the order of $h^m$, that is, $-O(h^m)<\epsilon <O(h^m)$. 
If $h$ is not so small, we have to deal with the equation (\ref{1}) 
directly. The equation (\ref{1}) is known as Mathieu's equation and 
the properties of this equation are investigated in detail.\cite{WW} 
In the following investigation, $h$ is actually small fortunately. Thus, it 
is not necessary to take care of the full treatment of this Mathieu's 
equation. 
Next, let us consider 
the equation of motion which describes the forced oscillation :
\begin{equation}\label{4}
{\ddot x}+\omega_0^2 x=F\cos \Omega t \ .
\end{equation}
The solution with initial conditions $x(0)={\dot x}(0)=0$ is obtained as 
\begin{equation}\label{5}
x(t)=F\cdot\frac{\cos \Omega t -\cos \omega_0 t}{\omega_0^2-\Omega^2}
=\frac{2F}{(\omega_0-\Omega)(\omega_0+\Omega)}\sin\frac{\omega_0-\Omega}{2}t
\cdot\sin\frac{\omega_0+\Omega}{2}t \ .
\end{equation}
As is well known, when $\Omega$ has a value near $\omega_0$, 
the behavior of the above solution shows the beat. 
Especially, $\Omega\approx \omega_0$, the amplitude of the solution 
becomes large linearly in time : 
\begin{equation}\label{6}
x(t)\approx ({F}/{2\omega_0})\cdot t \sin \omega_0t
\end{equation}

Now let us return to the main subject, namely, 
let us treat the $O(4)$-linear sigma model in the squeezed state 
approach. We start from the following Hamiltonian density of 
$O(4)$-linear sigma model: 
\begin{equation}\label{7}
    {\cal H} =
      \frac{1}{2} \pi_{a}({\mib x})^2 
      +\frac{1}{2} \nabla \phi_{a}({\mib x}) \cdot\nabla\phi_{a}({\mib x})
      +\lambda\left(\phi_{a}({\mib x})^2-{m^2}/{4\lambda} \right)^2 
         -H\phi_{0}({\mib x}) \ .
\end{equation}
The squeezed state is adopted as a trial state in the time-dependent 
variational principle : 
\begin{eqnarray}\label{8}
\ket{\Phi(t)}
=N\exp\left(i({\overline \pi}\cdot\phi-{\overline\varphi}\cdot\pi)\right)
\exp(\phi\cdot[-(G^{-1}-G^{(0)}{}^{-1})/4+i\Sigma]\cdot\phi)\ket{0} \ , \quad
\end{eqnarray}
where $N$ represents a normalization factor and 
${\overline \pi}\cdot\phi=\sum_{a=0}^3\int\! d^3\!{\mib x}
{\overline \pi}_a({\mib x},t)\phi_a({\mib x})$ and 
$\phi\cdot G^{-1}\cdot\phi=\sum_{a=0}^3\int\!\int\! d^3\!{\mib x}d^3\!{\mib y}
\phi_a({\mib x})G_{a}^{-1}({\mib x},{\mib y},t)\phi_a({\mib y})$ 
and so on. Here, $G^{(0)}_a=\bra{0}\phi_a({\mib x})\phi_a({\mib y})\ket{0}$. 
The expectation values of the field operators are obtained, 
for example, as 
\begin{eqnarray}\label{9}
& &
\bra{\Phi(t)}\phi_a({\mib x})\ket{\Phi(t)}
={\overline \varphi}_a({\mib x},t) \ , 
\nonumber\\
& &
\bra{\Phi(t)}\phi_a({\mib x})\phi_a({\mib y})\ket{\Phi(t)}
={\overline \varphi}_a({\mib x},t){\overline \varphi}_a({\mib y},t)
+G_a({\mib x},{\mib y},t) \ .
\end{eqnarray}
${\overline \varphi}_a({\mib x},y)$ and $G_a({\mib x},{\mib y},t)$ 
represent the mean field and the fluctuation around it, respectively. 
The time-dependence of these functions, together with ${\overline \pi}_a$ and 
$\Sigma_a$, are determined in the time-dependent variational principle :
\begin{equation}\label{10}
\delta\int_{t_1}^{t_2}dt \bra{\Phi(t)}i\frac{\partial}{\partial t}
-{\hat H}\ket{\Phi(t)}=0 \ , 
\end{equation}
where ${\hat H}=\int\!d^3\!{\mib x}{\cal H}$.
The two-point function $G_a({\mib x},{\mib y},t)$ can be further expanded 
in terms of the mode functions under the assumption of the translational 
invariance :
\begin{equation}\label{11}
G_{a}({\mib x},{\mib y},t)
=\int_{\bf k}
e^{i{\bf k}\cdot({\bf x}-{\bf y})}
\eta_{\bf k}^{a}(t)^2 \ , 
\end{equation}
where $\int_{\bf k}=\int d^3{\bf k}/{(2\pi)^3}$. 
The equations of motion for the condensate ${\overline \varphi}_a(t)$ 
and the fluctuation modes 
$\eta_{\bf k}^a(t)$ under the translational invariance can be 
expressed as 
\begin{eqnarray}\label{12}
  & &{\ddot {\overline \varphi}}_{a}(t)
     -m^2 {\overline \varphi}_{a}(t) 
     + 4\lambda {\overline \varphi}_{a}(t)^3
     +12\lambda \int_{\bf k}\eta_{\bf k}^{a}(t)^2\cdot 
       {\overline \varphi}_{a}(t) \nonumber\\
  & & \qquad\qquad\qquad\qquad
     +4\lambda \sum_{b\neq a}\left({\overline \varphi}_{b}(t)^2 
     +\int_{\bf k}\eta_{\bf k}^{b}(t)^2 \right)
     {\overline \varphi}_{a}(t) -H\delta_{a0} =0 \ , \nonumber\\
  & & {\ddot \eta}_{\bf k}^{a}(t) 
    + \biggl[{\mib k}^2-m^2+12\lambda {\overline \varphi}_{a}(t)^2 
    + 12\lambda\int_{{\bf k}'}\eta_{{\bf k}'}^{a}(t)^2 
       \nonumber\\
  & & \qquad\qquad\qquad\qquad
     +4\lambda \sum_{b\neq a}\left({\overline \varphi}_{b}(t)^2 
     +\int_{{\bf k}'}\eta_{{\bf k}'}^{b}(t)^2 
     \right) 
       \biggl]
    \eta_{\bf k}^{a}(t) 
    -{1}/{4 \eta_{\bf k}^{a}(t)^3} = 0 \ . \qquad
\end{eqnarray}
The derivation and the physical viewpoint in this approach are 
given in detail in Ref.\citen{TKI}. 
Hereafter, let us assume that the chiral condensate points in the sigma 
direction, that is, 
${\overline \varphi}_0\neq 0$ and ${\overline \varphi}_i=0$ for 
$i=1\sim 3$. Also, fluctuation modes of the pi directions are identical 
one another, which are denoted as $\eta_{\bf k}^{\pi}(t)$, that is, 
$\eta_{\bf k}^1=\eta_{\bf k}^2=\eta_{\bf k}^3\equiv \eta_{\bf k}^{\pi}$.
When the explicit chiral symmetry breaking term, $H$, is small, 
the static solutions in Eq.(\ref{12}) are given as 
\begin{equation}\label{13}
{\overline \varphi}_0\equiv 
\varphi_0=\sqrt{{M_\sigma^2}/{8\lambda}}-{H}/{2M_\sigma^2} \ ,
\quad
{\eta_{\bf k}^0}^2={1}/{2\sqrt{{\mib k}^2+M_\sigma^2}} \ , 
\quad
{\eta_{\bf k}^{\pi}}^2={1}/{2\sqrt{{\mib k}^2+M_\pi^2}} \ , 
\end{equation}
where the sigma meson mass $M_\sigma$ and the pion mass $M_\pi$ are 
defined as 
\begin{eqnarray}\label{14}
  & & M_{\sigma}^2
  =-m^2+12\lambda {\varphi}_{0}^2 
    + 12\lambda\int_{{\bf k}'}{\eta_{{\bf k}'}^{0}}^2 
     +12\lambda \int_{{\bf k}'}{\eta_{{\bf k}'}^{\pi}}^2 \ , \nonumber\\
& &M_{\pi}^2
    =-m^2+4\lambda {\varphi}_{0}^2 
    + 4\lambda\int_{{\bf k}'}{\eta_{{\bf k}'}^{0}}^2 
     +20\lambda \int_{{\bf k}'}{\eta_{{\bf k}'}^{\pi}}^2 \ .
\end{eqnarray}

Let us examine the time-dependent solutions around the static 
configurations. In Eq.(\ref{12}), the time-dependent variables 
can be expanded around the static solutions of Eq.(\ref{13}) : 
\begin{eqnarray}\label{15}
{\overline \varphi}_0(t)=\varphi_0+\delta\varphi(t) \ , \qquad
\eta_{\bf k}^a(t)=\eta_{\bf k}^a+\delta\eta_{\bf k}^a(t) \ . 
\end{eqnarray}
Here, we consider the late time of the chiral phase transition. 
Thus, $\delta\varphi(t)$ is small compared with $\varphi_0$. 
Further, we assume that $|\delta\eta_{\bf k}^a(t)|\ll |\eta_{\bf k}^a|$ : 
\begin{eqnarray}\label{16}
& &{| \delta\varphi(t) |}/{\varphi_0} \ll 1 \ , \qquad
|\delta\eta_{\bf k}^a(t)/\eta_{\bf k}^a| \ll 1 \ . 
\end{eqnarray}
From Eq.(\ref{9}), it is seen that the two-point function $G_a$ 
represents quantum fluctuations around the mean field. Namely, $G_a$ 
is small compared with $\varphi_0^2$. Thus, from Eq.(\ref{11}), 
we conclude that the following relation is satisfied : 
\begin{eqnarray}\label{17}
& &\varphi_0^2 \gg \int_{\bf k}\eta_{\bf k}^{a2} \ , \quad {\rm then}\ , 
\quad 
\left| \varphi_0\delta\varphi(t) \right| \gg 
\left| 
\int_{\bf k}\eta_{\bf k}^a \delta\eta_{\bf k}^a(t) 
\right| \ . 
\end{eqnarray}
Under the above approximations and the small explicit chiral symmetry 
breaking, the equations of motion for $\delta\varphi(t)$ and 
$\delta\eta_{\bf k}^a(t)$ are obtained from (\ref{12}) and (\ref{15}) as 
\begin{eqnarray}
& &\delta{\ddot \varphi}(t)+M_\sigma^2 \delta\varphi(t) = 0 \ , 
\label{18}\\
& &\delta{\ddot \eta}_{\bf k}^{\sigma}(t)
+\left[4({\mib k}^2+M_\sigma^2)+24\lambda\varphi_0 \delta\varphi(t)\right]
\delta\eta_{\bf k}^\sigma(t) 
= -24\lambda\eta_{\bf k}^{\sigma} \varphi_0 \delta\varphi(t) \ , 
\nonumber\\
& &\delta{\ddot \eta}_{\bf k}^{\pi}(t)
+\left[4({\mib k}^2+M_\pi^2)+8\lambda\varphi_0 \delta\varphi(t)\right]
\delta\eta_{\bf k}^\pi(t) 
= -8\lambda\eta_{\bf k}^{\pi} \varphi_0 \delta\varphi(t) \ , 
\label{19}
\end{eqnarray}
where $\eta_{\bf k}^0$ and $\delta\eta_{\bf k}^0(t)$ have been 
rewritten as $\eta_{\bf k}^\sigma$ and $\delta\eta_{\bf k}^\sigma(t)$, 
respectively. 
The equation (\ref{18}) is easily solved by using the sigma meson mass :
\begin{equation}\label{20}
\delta\varphi(t) =\delta\sigma \cos(M_\sigma t +\alpha) \ , 
\end{equation}
where $\delta\sigma$ and $\alpha$ are determined by the initial conditions. 
Here, we adopt $\alpha=\pi$ for simplicity. 
We further define the dimensionless variables : 
\begin{eqnarray}
& &\delta{\wtilde \eta}_{\bf k}^a(\tau)
=\sqrt{2M_a}\ \delta\eta_{\bf k}^a(t) \ , 
\qquad \tau={M_\sigma t}/{2} \ , 
\label{21}\\
& &
\omega_\sigma={4\sqrt{{\bf k}^2+M_\sigma^2}}/{M_\sigma} \ , \qquad\quad\qquad
\omega_\pi={4\sqrt{{\bf k}^2+M_\pi^2}}/{M_\sigma} \ , \nonumber\\
& &h_\sigma=6{\lambda\varphi_0^2}/({{\bf k}^2+M_\sigma^2})\cdot
{\delta\sigma}/{\varphi_0} \ , \qquad
h_\pi=2{\lambda\varphi_0^2}/({{\bf k}^2+M_\pi^2})\cdot
{\delta\sigma}/{\varphi_0} \ , \nonumber\\
& &F_\sigma=96\frac{\lambda\varphi_0^2}{M_\sigma^2}\cdot\!\!
\left(\frac{M_\sigma^2}{{\bf k}^2+M_\sigma^2}\right)^{1/4}\!\!\cdot
\frac{\delta\sigma}{\varphi_0} \ , \quad
F_\pi=32\frac{\lambda\varphi_0^2}{M_\sigma^2}\cdot\!\!
\left(\frac{M_\pi^2}{{\bf k}^2+M_\pi^2}\right)^{1/4}\!\!\cdot
\frac{\delta\sigma}{\varphi_0} \ . \qquad
\label{22}
\end{eqnarray}
By introducing the above dimensionless variables, the solution (\ref{20}) and 
the equations of motion (\ref{19}) are recast into the simple forms as 
\begin{eqnarray}
& &\delta\varphi(\tau)=-\delta\sigma\cos\gamma\tau \ , \qquad
\gamma=2 \ , 
\label{23}\\
& &\left(\frac{d^2}{d\tau^2}+\omega_\sigma^2
[1-h_\sigma\cos\gamma\tau]\right)\delta{\wtilde \eta}_{\bf k}^\sigma(\tau)
=F_\sigma\cos\gamma\tau \ , 
\label{24}\\
& &\left(\frac{d^2}{d\tau^2}+\omega_\pi^2
[1-h_\pi\cos\gamma\tau]\right)\delta{\wtilde \eta}_{\bf k}^\pi(\tau)
=F_\pi\cos\gamma\tau \ .
\label{25}
\end{eqnarray}
If $F_\sigma$ ($F_\pi$) is negligible, then the equation 
(\ref{24}) ((\ref{25})) is reduced to Eq.(\ref{1}). In this case, 
the existence of the unstable solutions for 
$\delta{\wtilde \eta}_{\bf k}^a(\tau)$ may be expected. 
On the other hand, if $h_\sigma$ ($h_\pi$) is negligible or even if 
the parameters do not offer the unstable regions, the forced oscillation 
may be realized by the effect of $F_\sigma$ ($F_\pi$). 

In the previous paper,\cite{TKI} 
the behavior of the time-dependent solutions in Eq.(\ref{12}) was 
investigated numerically. 
It was shown that the amplification of the amplitude of the pion modes 
with low momenta was realized even in the late time. 
However, it was seen that 
the amplitude of the second excited mode was also amplified 
(See, Fig.3 in Ref.\citen{TKI}). We can now understand their behaviors 
by using Eqs.(\ref{24}) and (\ref{25}) approximately. 
The model parameters $m$, $\lambda$ and $H$ are adopted so as to reproduce 
the pion mass, the sigma meson mass and the pion decay constant 
in the static case in Eqs.(\ref{13}) and (\ref{14}). 
The box normalization with the spatial 
length $L$ ($=10$ fm) is applied, in which the periodic boundary conditions 
for the fluctuation modes are imposed. Thus, the allowed values of momenta 
are restricted as follows : $k_x=(2\pi/L)n_x$ and so on, where $n_x$ 
is integer. 
The three momentum cutoff is also applied to regularize the momentum 
integration. The fluctuation modes are taken into account up to 
$n^2=n_x^2+n_y^2+n_z^2 \leq 8^2$ corresponding to the three momentum cutoff 
about 1 GeV (990 MeV). 
The values of variables given in Eq.(\ref{22}) are collected up to 
$n=4$ in Table I., where $\delta=\delta\sigma/\varphi_0\approx O(10^{-1})$ 
in the late time of the time-evolution of the chiral condensate.

\begin{table}[t]
\caption{The values of dimensionless variables based on the parameters 
used in the previous paper.\cite{TKI} Here, 
$\delta=\delta\sigma/\varphi_0$}
\label{table:1}
\begin{center}
\begin{tabular}{c|cccccc} \hline \hline
mode & $\omega_\sigma$ & $\omega_\pi$ & $ h_\sigma$ & $h_\pi$ & 
$F_\sigma$ & $F_\pi$ \\ \hline 
$n=0$ & 4.00 & 0.92 & 0.766$\times\delta$ & 4.83$\times\delta$ 
& 12.3$\times\delta$ & 4.09$\times\delta$ \\
$n=1$ & 4.08 & 1.24 & 0.735$\times\delta$ & 2.67$\times\delta$ 
& 12.1$\times\delta$ & 3.52$\times\delta$ \\ 
$n=2$ & 4.33 & 1.89 & 0.655$\times\delta$ & 1.14$\times\delta$ 
& 11.8$\times\delta$ & 2.79$\times\delta$ \\
$n=3$ & 4.70 & 2.64 & 0.554$\times\delta$ & 0.586$\times\delta$ 
& 11.3$\times\delta$ & 2.41$\times\delta$ \\
$n=4$ & 5.19 & 3.43 & 0.456$\times\delta$ & 0.348$\times\delta$ 
& 10.8$\times\delta$ & 2.12$\times\delta$ \\
\hline
\end{tabular}
\end{center}
\end{table}

First, let us consider the pion modes. It should be noted that $\gamma=2$. 
For the lowest pion mode with $n=0$, $\omega_\pi=0.92$. Therefore, 
it would not 
expected that the amplification due to the forced oscillation occurs. 
We thus investigate the possibility of the parametric resonance by 
setting $F_\pi\approx 0$ safely. By noting $2\omega_\pi \approx 2$, 
let us examine the parametric resonance with $m=1$ in the case Eq.(\ref{2}). 
The unstable region is given as 
\begin{equation}\label{26}
-h_\pi \omega_\pi/2 < \epsilon < h_\pi\omega_\pi/2 \  \qquad 
\hbox{\rm with}\ \ 
\gamma=2=2\omega_\pi+\epsilon \ , \quad
(\epsilon \approx 0.16 )\ .
\end{equation}
The condition (\ref{26}) reads together with Table I. as 
\begin{equation}\label{27}
\delta\equiv \delta\sigma/\varphi_0 > 0.072 \ . 
\end{equation}
Thus, in the above region for $\delta\sigma$, the parametric resonance 
occurs and the amplitude of the lowest pion mode is amplified. 
This behavior appears in Fig.3 in Ref.\citen{TKI}. 
Similarly, let us consider the pion mode with $n=1$, the first excited mode. 
As is similar to the case $n=0$, it is not necessary to take care of 
the amplification due to the forced oscillation because 
$\omega_\pi=1.24$. For $m=1$, the same condition as (\ref{26}) should be 
satisfied to realize the parametric resonance, except for the value 
of $\epsilon$ : $\epsilon=2(1-\omega_\pi)\approx -0.48$. 
Thus, the unstable region is rewritten as 
\begin{equation}\label{28}
\delta\equiv \delta\sigma/\varphi_0 > 0.29 \ . 
\end{equation}
The parametric resonance is then realized. 
For $m=2$, the unstable region is given as 
\begin{equation}\label{29}
-(5/24)\cdot h_\pi^2 \omega_\pi < \epsilon < (1/24)\cdot h_\pi^2\omega_\pi 
\qquad {\rm with}\ \ \gamma=2=\omega_\pi+\epsilon \, \quad 
(\epsilon\approx 0.76) \ .
\end{equation}
However, $(1/24)\cdot h_\pi^2\omega_\pi\approx 0.37 
(\delta\sigma/\varphi_0)^2$ is obtained in this parameterization. 
Therefore, the parametric resonance with $m=2$ does not occur. 

Secondly, let us investigate the pion mode with $n=2$, 
the second excited mode. If the parametric amplification occur,  
the case $m=2$ is realized because $\omega_\pi=1.89$. This condition is 
given in Eq.(\ref{29}) except for $\epsilon$ : $\epsilon \approx 0.11$. 
However, the upper limit $(1/24)\cdot h_\pi^2\omega_\pi$ gives a severe 
limit : $0.10\cdot(\delta\sigma/\varphi_0)^2$. Thus, the condition (\ref{29}) 
can not be satisfied. 
On the other hand, it is possible that the amplitude of the pion mode 
with $n=2$ is amplified by the mechanism of the forced oscillation. 
The value of $\omega_\pi\ (=1.89)$ is near $\gamma\ (=2)$. 
If we neglect $h_\pi$, 
the equation (\ref{25}) is same as (\ref{4}) and the solution 
is given in Eq.({\ref{6}). Thus, we obtain 
\begin{equation}\label{30}
\delta{\wtilde \eta}_{n=2}^{\pi}(\tau)
\approx 0.71 (\delta\sigma/\varphi_0) \tau \sin \omega_\pi \tau \ .
\end{equation}
Thus, the amplitude increases linearly in time. 
This behavior is seen in Fig.3 in Ref.\citen{TKI}.

Finally, let us consider the sigma mode. 
The amplification due to the forced oscillation is impossible 
because the frequency $\omega_\sigma$ is always greater than 2 $(=\gamma)$. 
Here, let us investigate the possibility of the parametric amplification. 
The lowest mode with $n=0$ is the most powerful candidate. 
Since $\omega_\sigma=4$, the case $m=4$ is applicable : 
$\gamma=2=\omega_\sigma/2+\epsilon$, where $\epsilon=0$. 
The unstable region is given as $-O(h_\sigma^4)<\epsilon<O(h_\sigma^4)$ 
for $m=4$ case. Since $\epsilon=0$, the parametric amplification occurs. 
However, the unstable solution is proportional to 
$e^{s\tau}$, in which $s$ is of the order of $h_\sigma^4$ for $m=4$ case. 
From Table I., $h_\sigma^4$ is about $3.4\times 10^{-5}$ with 
$\delta\sigma/\varphi_0=10^{-1}$. This is very small value. Thus, 
the amplification occurs slowly in time. In the time region 
under consideration, the amplification may not be seen explicitly. 
In the sigma modes with higher momenta, the parametric resonance would not 
occur because the unstable window is very narrow.

It should be noted here that the mechanism amplifying the amplitudes of 
fluctuation modes, i.e., the parametric resonance or the forced oscillation, 
depends on the model parameters. 
However, in conclusion, it should be pointed out 
that there is a possibility of the amplification due to 
the forced oscillation as well as the parametric resonance. 
In this paper, the parametric resonance and the forced oscillation 
are treated separately in Eqs.(\ref{24}) and (\ref{25}). 
It is interesting to investigate both the mechanisms in the unified way. 
The important ingredients may be the properties of Mathieu's equation 
and the Mathieu function.

\vspace{0.4cm}

This work was supported by Grant-in-Aid 
for Scientific Research from the Ministry of Education, Culture, 
Sports, Science and Technology (No.13740159).


\begin{thebibliography}{99}
\bibitem{TKI}
Y.Tsue, A. Koike and N. Ikezi, 
        Prog.~Theor.~Phys. {\bf 106} (2001), 807. 
\bibitem{MM}
S. Mr\'owczy\'nski and B. M\"uller, 
        Phys.~Lett.~{\bf B363} (1995), 1.  
\bibitem{HM}
H. Hiro-Oka and H. Minakata, 
        Phys.~Rev.~{\bf C61} (2000), 044903. 
\bibitem{M}
S. Maedan, 
        Phys. Lett. {\bf B512} (2001), 73. 
\bibitem{I}
M. Ishihara, 
        Phys. Rev. {\bf C62} (2000), 054908.
\bibitem{LL}
L. D. Landau and E. M. Lifshitz,         
        {\it Mechanics}, 
        (Butterworth Heinemann, 1976), \S 27.
\bibitem{WW}
E. T. Whittaker and G. N. Watson, 
        {\it A course of Modern Analysis}, 
        (Cambridge at the University Press, 1935), p.404.
\end{thebibliography}
\end{document}